\documentstyle[12pt]{article}
\begin{document}
\begin{center}
{\Large\bf Nucleation in two dimensional Ising system: Classical theory
and computer simulation}
\end{center}
\vspace {0.4 cm}
\begin{center} {\large Muktish Acharyya$^*$} 
\end{center}
\vspace {0.3 cm}
\begin{center} {\it Institut for Theoretical Physics,} \end{center}
\begin{center} {\it Cologne University, 50923 K\"oln, Germany} \end{center}
\vspace {1.0 cm}
\noindent {\bf Abstract:} We have studied the nucleation in 
the two dimensional Ising model
by Monte Carlo simulation. The nucleation time has been studied as a 
function of the magnetic field for various system sizes. The logarithm of
the nucleation time is found to be proportional to the inverse of the magnetic
field.  
The crossover field and time are studied as a function of system size.
The observed variations are consistent with the prediction of classical
nucleation theory.

\vspace {2 cm}
\leftline {\bf Keywords: Ising model, Nucleation, Monte Carlo simulation }
\leftline {\bf PACS Numbers: 05.50 +q}
\vspace {4 cm}
\leftline {----------------------------}
\leftline {\bf $^*$E-mail: muktish@thp.uni-koeln.de}
\newpage

\leftline {\bf I. Introduction}

The dynamical aspects of Ising model is an active area of modern research.
How does the magnetisation relax towards its equilibrium value, if we start
the dynamics with all spins parallel ? How long is the lifetime of a 
metastable state if a magnetic field is antiparallel to the initial spin
orientation ? Can one answer all these questions in the light of growing
and shrinking droplets ?

What happens if all spins are up in presence of a small opposite magnetic
field and the system is below its critical temperature ($T_c$) ? The 
magnetisation first settles to a metastable state, and then a droplet
larger than a critical size is formed. As the time passes, this droplet
grows radially and the magnetisation jumps to a negative value. Classical
nucleation theory (CNT) \cite{bd} predicts that the logarithm of the
nucleation rate (number of supercritical droplets formed per unit time
per unit volume) to be asymptotically proportional to $h^{1-d}$ in $d$
dimension, where $h$ is the normalized magnetic field. This has been
verified in the three dimensional Ising model by Monte Carlo simulation
\cite{ds}. There are some difficuties in measuring the nucleation rates
by checking how long the magnetisation takes to leave its metastable
value. In the asymptotic limit of field ($h$) going to zero for a finite
lattice size, only one supercritical droplet will be formed and it grows
to cover the whole lattice. This is the proper nucleation regime and the
nucleation rate is the reciprocal of the product of nucleation time and
lattice volume. On the other hand, in the coalescence regime, 
with the lattice size
going to infinity at fixed field ($h$), many such supercritical droplets
will be formed at a time and they grow and coalesce and as a
consequence the magnetisation switches sign. 
This effect, already discussed by Binder and M\"uller-
Krumbhaar \cite{mul} and mathematically shown by Schonmann \cite{sch}, was
demonstrated by Ray and Wang \cite{rw} for Swendsen-Wang dynamics.

In this paper, we have studied the nucleation in two dimensional Ising system by
Monte Carlo simulation with heat-bath dynamics. We have also studied the system 
size dependent crossover (from nucleation regime to coalescence regime)
and compared the simulational results with the results of classical
nucleation theory.
 
\bigskip
\leftline {\bf II. Classical nucleation theory}

We review briefly the results of classical nucleation
theory far below $T_c$. The equilibrium number 
(per site) $n_s$ of droplets, containing $s$
spins is
$$n_s \sim \exp(-F_s/KT) $$
\noindent where $F_s$ is the formation free energy of the droplet of size $s$
and $K$ is the Boltzmann constant.
CNT assumes a spherical droplet shape and takes
$$F_s/KT = -h s + 2 \pi^{1/2} s^{1/2}\sigma$$ 
\noindent where $\sigma$ is the surface tension. The critical
size $s^*$ of a droplet which maximises the free energy, is 
$$ s^* = \pi \sigma^2/h^2$$
\noindent and
$$F_{max} = \pi \sigma^2/h.$$
\noindent The number $n_{s^*}$ of droplets having the critical size is
$$n_{s^*} \sim \exp(-F_{max}/KT) \sim \exp(-\pi \sigma^2/hKT).$$
\noindent The nucleation rate ($J$) is proportinal to $n_{s^*}$
$$J \sim n_{s^*} \sim \exp(-\pi \sigma^2/hKT).$$
\noindent In the nucleation regime, the nucleation time $\tau$ is inversely
proportinal to the nucleation rate $J$, 
$$\tau \sim J^{-1} \sim \exp(\pi \sigma^2/hKT). \eqno(1)$$
\noindent In the coalescence regime, the number of spins in
 a supercritical droplet
will grow as $t^d$ ($d = 2$ here). For a steady rate of nucleation, the 
rate of change
of magnetisation is 
$Jt^d$, for a fixed change ($\Delta m$) in magnetisation during the
nucleation time $\tau$, 
$$\Delta m \sim \int_0^{\tau} J 
t^d dt \sim (J \tau)^{d+1}.$$ 
\noindent So, in the
coalescence regime, 
$$\tau \sim J^{-1/(d+1)} \sim \exp(\pi \sigma^2/3hKT). \eqno(2)$$

\bigskip
\leftline {\bf III. Numerical Results}

We have considered a square lattice (of linear size $L$)
with nearest neighbour ferromagnetic interaction.
Initially, all spins are
taken parallel (up) and used the heat-bath dynamics to simulate the
nucleation phenomena. 
We have now allowed the system to relax to another equilibrium state via
a metastable state in presence of an opposite (negative) magnetic field
at a temperature ($t = J/KT = 5/8$) below the critical temperature ($J/KT_c =
0.440687$).
We have used helical boundary condition in one direction
and periodic boundary condition in the other direction.

In the nucleation regime, the true nucleation time is quite large and
fluctuates enormously. Here, we define the nucleation time as the time required
by the system to have the magnetisation below 0.9. This choice is quite 
arbitrary and the results do not depend considerably on the choice of this
threshold. 
The range of temperature has been chosen in such a way that the metastable
values, at those temperatures, always lie above 0.9.
Due to the huge fluctuations in the nucleation time, to avoid the
waste of computaional time we have taken the {\it median}
 nucleation time instead of
taking the (algebraic) mean.

Figure 1  shows the plot of nucleation time (at a fixed temperature,
$J/KT = 5/8$) versus the normalized magnetisation $h$ for different 
system sizes. It shows clearly the crossover from a nucleation regime
(small lattices, smaller fields, long nucleation time) due to a single
supercritical droplet to a coalescence regime with many such supercritical
droplets (large lattices, larger fields, short nucleation time). The
larger the lattice is, the longer is the nucleation time above which the
proper nucleation regime can be found. So, by taking a smaller lattice,
it is convenient to observe the crossover from nucleation to a coalescence
regime, within shorter period
(simulational time scale). Numerically, we found that the 
logarithm of the nucleation time
is proportional to the inverse field ($1/h$) which is consistent with
the results obtained from CNT
(eqn (1)). The slopes are also been estimated numerically.
The slope in the nucleation regime is nearly 0.79 and that in the coalescence
regime is approximately 0.27, which is roughly 1/3 of that in the nucleation
regime. This is consistent with the results from CNT. We know \cite{ff}
$$a \sigma(T)/KT_c = -2 (T/T_c -1) \log_e (\sqrt 2 +1)$$
\noindent where $a$ is the lattice separation ($a = 1$, here).
In the above formula, putting $J/KT = 5/8$, we get
$$\pi \sigma^2 (T)/T = 2.733.$$
\noindent So, according to CNT,
$$\tau \sim \exp(2.733/h) \sim 10^{(1.187/h)}.$$
\noindent So, the slope
(at $J/KT = 5/8$)
in the nucleation regime, is estimated to be equal to 1.187 \cite{ff}. 
However, the simulation results give 0.79.
This
mismatch may be due to the nonspherical 
shape of the droplets (assumed in CNT).	 
It would be worthmentioning here that the theoretically calculated 
\cite{ff} slope
(using the diagonal interface tension) at the temperature used is 1.900.

We have also studied the variation of the crossover field ($h_c$; at which
the crossover from nucleation regime to a coalescence regime starts) and 
the crossover time ($\tau_c$; the nucleation time at the 
crossover point) with the system
size. From CNT, the nucleation time and the field is related as
$$\tau \sim {1 \over {J \times L^2}} \sim \exp(\pi \sigma^2/hKT)/L^2$$
\noindent At the crossover point ($\tau_c, h_c$)
$$\tau_c \sim \exp(\pi \sigma^2/h_c KT)/L^2.$$
\noindent If we assume $\tau_c \sim L^2$, then
$$h_c \sim 1/\log(L). \eqno(3)$$
\noindent Figure 2 shows that plot of crossover field ($h_c$) versus 
$1/\log(L)$ and it is indeed a straight line, which is consistent with
the resuts of CNT (eqn (3)). 
Figure 3 shows the variation of crossover time ($\tau_c$)
with respect to $L^2$, which is a straight line. This also proves that the
assumption $\tau_c \sim L^2$ is correct in the derivation of equation (3).

\bigskip
\leftline {\bf IV. Summary}

We have studied the nucleation in the two dimensional Ising system by
Monte Carlo simulation with heat-bath dynamics. The logarithm of
the nucleation time is found to be proportional to the inverse magnetic field.
The proportionality constant (related with the surface tension) has been
estimated and found to be close to that calculated from CNT. The nucleation
time has been studied for various system sizes which shows a crossover from
the nucleation regime to a coalescence regime. The crossover field is
found to be
proportional to the inverse of the logarithm of the system size and the
crossover time is 
found to be proportional to the square of the system size. 
The results are consistent
with the prediction of classical nucleation theory.

\bigskip
\noindent {\bf Acknowledgements:} Sonderforschungbereich 
341 is gratefully acknowledged for financial support. Author would like to
thank D. Stauffer for suggesting the problem and useful comments.

\newpage
%FIG 1
\setlength{\unitlength}{0.240900pt}
\ifx\plotpoint\undefined\newsavebox{\plotpoint}\fi
\sbox{\plotpoint}{\rule[-0.200pt]{0.400pt}{0.400pt}}%
\begin{picture}(1500,900)(0,0)
\font\gnuplot=cmr10 at 10pt
\gnuplot
\sbox{\plotpoint}{\rule[-0.200pt]{0.400pt}{0.400pt}}%
\put(220.0,113.0){\rule[-0.200pt]{0.400pt}{184.048pt}}
\put(220.0,113.0){\rule[-0.200pt]{4.818pt}{0.400pt}}
\put(198,113){\makebox(0,0)[r]{1}}
\put(1416.0,113.0){\rule[-0.200pt]{4.818pt}{0.400pt}}
\put(220.0,159.0){\rule[-0.200pt]{2.409pt}{0.400pt}}
\put(1426.0,159.0){\rule[-0.200pt]{2.409pt}{0.400pt}}
\put(220.0,220.0){\rule[-0.200pt]{2.409pt}{0.400pt}}
\put(1426.0,220.0){\rule[-0.200pt]{2.409pt}{0.400pt}}
\put(220.0,251.0){\rule[-0.200pt]{2.409pt}{0.400pt}}
\put(1426.0,251.0){\rule[-0.200pt]{2.409pt}{0.400pt}}
\put(220.0,266.0){\rule[-0.200pt]{4.818pt}{0.400pt}}
\put(198,266){\makebox(0,0)[r]{10}}
\put(1416.0,266.0){\rule[-0.200pt]{4.818pt}{0.400pt}}
\put(220.0,312.0){\rule[-0.200pt]{2.409pt}{0.400pt}}
\put(1426.0,312.0){\rule[-0.200pt]{2.409pt}{0.400pt}}
\put(220.0,373.0){\rule[-0.200pt]{2.409pt}{0.400pt}}
\put(1426.0,373.0){\rule[-0.200pt]{2.409pt}{0.400pt}}
\put(220.0,404.0){\rule[-0.200pt]{2.409pt}{0.400pt}}
\put(1426.0,404.0){\rule[-0.200pt]{2.409pt}{0.400pt}}
\put(220.0,419.0){\rule[-0.200pt]{4.818pt}{0.400pt}}
\put(198,419){\makebox(0,0)[r]{100}}
\put(1416.0,419.0){\rule[-0.200pt]{4.818pt}{0.400pt}}
\put(220.0,465.0){\rule[-0.200pt]{2.409pt}{0.400pt}}
\put(1426.0,465.0){\rule[-0.200pt]{2.409pt}{0.400pt}}
\put(220.0,525.0){\rule[-0.200pt]{2.409pt}{0.400pt}}
\put(1426.0,525.0){\rule[-0.200pt]{2.409pt}{0.400pt}}
\put(220.0,557.0){\rule[-0.200pt]{2.409pt}{0.400pt}}
\put(1426.0,557.0){\rule[-0.200pt]{2.409pt}{0.400pt}}
\put(220.0,571.0){\rule[-0.200pt]{4.818pt}{0.400pt}}
\put(198,571){\makebox(0,0)[r]{1000}}
\put(1416.0,571.0){\rule[-0.200pt]{4.818pt}{0.400pt}}
\put(220.0,617.0){\rule[-0.200pt]{2.409pt}{0.400pt}}
\put(1426.0,617.0){\rule[-0.200pt]{2.409pt}{0.400pt}}
\put(220.0,678.0){\rule[-0.200pt]{2.409pt}{0.400pt}}
\put(1426.0,678.0){\rule[-0.200pt]{2.409pt}{0.400pt}}
\put(220.0,709.0){\rule[-0.200pt]{2.409pt}{0.400pt}}
\put(1426.0,709.0){\rule[-0.200pt]{2.409pt}{0.400pt}}
\put(220.0,724.0){\rule[-0.200pt]{4.818pt}{0.400pt}}
\put(198,724){\makebox(0,0)[r]{10000}}
\put(1416.0,724.0){\rule[-0.200pt]{4.818pt}{0.400pt}}
\put(220.0,770.0){\rule[-0.200pt]{2.409pt}{0.400pt}}
\put(1426.0,770.0){\rule[-0.200pt]{2.409pt}{0.400pt}}
\put(220.0,831.0){\rule[-0.200pt]{2.409pt}{0.400pt}}
\put(1426.0,831.0){\rule[-0.200pt]{2.409pt}{0.400pt}}
\put(220.0,862.0){\rule[-0.200pt]{2.409pt}{0.400pt}}
\put(1426.0,862.0){\rule[-0.200pt]{2.409pt}{0.400pt}}
\put(220.0,877.0){\rule[-0.200pt]{4.818pt}{0.400pt}}
\put(198,877){\makebox(0,0)[r]{100000}}
\put(1416.0,877.0){\rule[-0.200pt]{4.818pt}{0.400pt}}
\put(220.0,113.0){\rule[-0.200pt]{0.400pt}{4.818pt}}
\put(220,68){\makebox(0,0){0}}
\put(220.0,857.0){\rule[-0.200pt]{0.400pt}{4.818pt}}
\put(463.0,113.0){\rule[-0.200pt]{0.400pt}{4.818pt}}
\put(463,68){\makebox(0,0){2}}
\put(463.0,857.0){\rule[-0.200pt]{0.400pt}{4.818pt}}
\put(706.0,113.0){\rule[-0.200pt]{0.400pt}{4.818pt}}
\put(706,68){\makebox(0,0){4}}
\put(706.0,857.0){\rule[-0.200pt]{0.400pt}{4.818pt}}
\put(950.0,113.0){\rule[-0.200pt]{0.400pt}{4.818pt}}
\put(950,68){\makebox(0,0){6}}
\put(950.0,857.0){\rule[-0.200pt]{0.400pt}{4.818pt}}
\put(1193.0,113.0){\rule[-0.200pt]{0.400pt}{4.818pt}}
\put(1193,68){\makebox(0,0){8}}
\put(1193.0,857.0){\rule[-0.200pt]{0.400pt}{4.818pt}}
\put(1436.0,113.0){\rule[-0.200pt]{0.400pt}{4.818pt}}
\put(1436,68){\makebox(0,0){10}}
\put(1436.0,857.0){\rule[-0.200pt]{0.400pt}{4.818pt}}
\put(220.0,113.0){\rule[-0.200pt]{292.934pt}{0.400pt}}
\put(1436.0,113.0){\rule[-0.200pt]{0.400pt}{184.048pt}}
\put(220.0,877.0){\rule[-0.200pt]{292.934pt}{0.400pt}}
\put(45,495){\makebox(0,0){$\tau$}}
\put(828,23){\makebox(0,0){$1/h$}}
\put(220.0,113.0){\rule[-0.200pt]{0.400pt}{184.048pt}}
\put(220,232){\usebox{\plotpoint}}
\multiput(220.00,232.60)(1.651,0.468){5}{\rule{1.300pt}{0.113pt}}
\multiput(220.00,231.17)(9.302,4.000){2}{\rule{0.650pt}{0.400pt}}
\multiput(232.00,236.60)(1.797,0.468){5}{\rule{1.400pt}{0.113pt}}
\multiput(232.00,235.17)(10.094,4.000){2}{\rule{0.700pt}{0.400pt}}
\multiput(245.00,240.60)(1.651,0.468){5}{\rule{1.300pt}{0.113pt}}
\multiput(245.00,239.17)(9.302,4.000){2}{\rule{0.650pt}{0.400pt}}
\multiput(257.00,244.59)(1.267,0.477){7}{\rule{1.060pt}{0.115pt}}
\multiput(257.00,243.17)(9.800,5.000){2}{\rule{0.530pt}{0.400pt}}
\multiput(269.00,249.60)(1.651,0.468){5}{\rule{1.300pt}{0.113pt}}
\multiput(269.00,248.17)(9.302,4.000){2}{\rule{0.650pt}{0.400pt}}
\multiput(281.00,253.60)(1.797,0.468){5}{\rule{1.400pt}{0.113pt}}
\multiput(281.00,252.17)(10.094,4.000){2}{\rule{0.700pt}{0.400pt}}
\multiput(294.00,257.60)(1.651,0.468){5}{\rule{1.300pt}{0.113pt}}
\multiput(294.00,256.17)(9.302,4.000){2}{\rule{0.650pt}{0.400pt}}
\multiput(306.00,261.60)(1.651,0.468){5}{\rule{1.300pt}{0.113pt}}
\multiput(306.00,260.17)(9.302,4.000){2}{\rule{0.650pt}{0.400pt}}
\multiput(318.00,265.60)(1.797,0.468){5}{\rule{1.400pt}{0.113pt}}
\multiput(318.00,264.17)(10.094,4.000){2}{\rule{0.700pt}{0.400pt}}
\multiput(331.00,269.59)(1.267,0.477){7}{\rule{1.060pt}{0.115pt}}
\multiput(331.00,268.17)(9.800,5.000){2}{\rule{0.530pt}{0.400pt}}
\multiput(343.00,274.60)(1.651,0.468){5}{\rule{1.300pt}{0.113pt}}
\multiput(343.00,273.17)(9.302,4.000){2}{\rule{0.650pt}{0.400pt}}
\multiput(355.00,278.60)(1.651,0.468){5}{\rule{1.300pt}{0.113pt}}
\multiput(355.00,277.17)(9.302,4.000){2}{\rule{0.650pt}{0.400pt}}
\multiput(367.00,282.60)(1.797,0.468){5}{\rule{1.400pt}{0.113pt}}
\multiput(367.00,281.17)(10.094,4.000){2}{\rule{0.700pt}{0.400pt}}
\multiput(380.00,286.60)(1.651,0.468){5}{\rule{1.300pt}{0.113pt}}
\multiput(380.00,285.17)(9.302,4.000){2}{\rule{0.650pt}{0.400pt}}
\multiput(392.00,290.60)(1.651,0.468){5}{\rule{1.300pt}{0.113pt}}
\multiput(392.00,289.17)(9.302,4.000){2}{\rule{0.650pt}{0.400pt}}
\multiput(404.00,294.59)(1.378,0.477){7}{\rule{1.140pt}{0.115pt}}
\multiput(404.00,293.17)(10.634,5.000){2}{\rule{0.570pt}{0.400pt}}
\multiput(417.00,299.60)(1.651,0.468){5}{\rule{1.300pt}{0.113pt}}
\multiput(417.00,298.17)(9.302,4.000){2}{\rule{0.650pt}{0.400pt}}
\multiput(429.00,303.60)(1.651,0.468){5}{\rule{1.300pt}{0.113pt}}
\multiput(429.00,302.17)(9.302,4.000){2}{\rule{0.650pt}{0.400pt}}
\multiput(441.00,307.60)(1.651,0.468){5}{\rule{1.300pt}{0.113pt}}
\multiput(441.00,306.17)(9.302,4.000){2}{\rule{0.650pt}{0.400pt}}
\multiput(453.00,311.60)(1.797,0.468){5}{\rule{1.400pt}{0.113pt}}
\multiput(453.00,310.17)(10.094,4.000){2}{\rule{0.700pt}{0.400pt}}
\multiput(466.00,315.60)(1.651,0.468){5}{\rule{1.300pt}{0.113pt}}
\multiput(466.00,314.17)(9.302,4.000){2}{\rule{0.650pt}{0.400pt}}
\multiput(478.00,319.59)(1.267,0.477){7}{\rule{1.060pt}{0.115pt}}
\multiput(478.00,318.17)(9.800,5.000){2}{\rule{0.530pt}{0.400pt}}
\multiput(490.00,324.60)(1.797,0.468){5}{\rule{1.400pt}{0.113pt}}
\multiput(490.00,323.17)(10.094,4.000){2}{\rule{0.700pt}{0.400pt}}
\multiput(503.00,328.60)(1.651,0.468){5}{\rule{1.300pt}{0.113pt}}
\multiput(503.00,327.17)(9.302,4.000){2}{\rule{0.650pt}{0.400pt}}
\multiput(515.00,332.60)(1.651,0.468){5}{\rule{1.300pt}{0.113pt}}
\multiput(515.00,331.17)(9.302,4.000){2}{\rule{0.650pt}{0.400pt}}
\multiput(527.00,336.60)(1.651,0.468){5}{\rule{1.300pt}{0.113pt}}
\multiput(527.00,335.17)(9.302,4.000){2}{\rule{0.650pt}{0.400pt}}
\multiput(539.00,340.60)(1.797,0.468){5}{\rule{1.400pt}{0.113pt}}
\multiput(539.00,339.17)(10.094,4.000){2}{\rule{0.700pt}{0.400pt}}
\multiput(552.00,344.59)(1.267,0.477){7}{\rule{1.060pt}{0.115pt}}
\multiput(552.00,343.17)(9.800,5.000){2}{\rule{0.530pt}{0.400pt}}
\multiput(564.00,349.60)(1.651,0.468){5}{\rule{1.300pt}{0.113pt}}
\multiput(564.00,348.17)(9.302,4.000){2}{\rule{0.650pt}{0.400pt}}
\multiput(576.00,353.60)(1.651,0.468){5}{\rule{1.300pt}{0.113pt}}
\multiput(576.00,352.17)(9.302,4.000){2}{\rule{0.650pt}{0.400pt}}
\multiput(588.00,357.60)(1.797,0.468){5}{\rule{1.400pt}{0.113pt}}
\multiput(588.00,356.17)(10.094,4.000){2}{\rule{0.700pt}{0.400pt}}
\multiput(601.00,361.60)(1.651,0.468){5}{\rule{1.300pt}{0.113pt}}
\multiput(601.00,360.17)(9.302,4.000){2}{\rule{0.650pt}{0.400pt}}
\multiput(613.00,365.60)(1.651,0.468){5}{\rule{1.300pt}{0.113pt}}
\multiput(613.00,364.17)(9.302,4.000){2}{\rule{0.650pt}{0.400pt}}
\multiput(625.00,369.59)(1.378,0.477){7}{\rule{1.140pt}{0.115pt}}
\multiput(625.00,368.17)(10.634,5.000){2}{\rule{0.570pt}{0.400pt}}
\multiput(638.00,374.60)(1.651,0.468){5}{\rule{1.300pt}{0.113pt}}
\multiput(638.00,373.17)(9.302,4.000){2}{\rule{0.650pt}{0.400pt}}
\multiput(650.00,378.60)(1.651,0.468){5}{\rule{1.300pt}{0.113pt}}
\multiput(650.00,377.17)(9.302,4.000){2}{\rule{0.650pt}{0.400pt}}
\multiput(662.00,382.60)(1.651,0.468){5}{\rule{1.300pt}{0.113pt}}
\multiput(662.00,381.17)(9.302,4.000){2}{\rule{0.650pt}{0.400pt}}
\multiput(674.00,386.60)(1.797,0.468){5}{\rule{1.400pt}{0.113pt}}
\multiput(674.00,385.17)(10.094,4.000){2}{\rule{0.700pt}{0.400pt}}
\multiput(687.00,390.60)(1.651,0.468){5}{\rule{1.300pt}{0.113pt}}
\multiput(687.00,389.17)(9.302,4.000){2}{\rule{0.650pt}{0.400pt}}
\multiput(699.00,394.59)(1.267,0.477){7}{\rule{1.060pt}{0.115pt}}
\multiput(699.00,393.17)(9.800,5.000){2}{\rule{0.530pt}{0.400pt}}
\multiput(711.00,399.60)(1.797,0.468){5}{\rule{1.400pt}{0.113pt}}
\multiput(711.00,398.17)(10.094,4.000){2}{\rule{0.700pt}{0.400pt}}
\multiput(724.00,403.60)(1.651,0.468){5}{\rule{1.300pt}{0.113pt}}
\multiput(724.00,402.17)(9.302,4.000){2}{\rule{0.650pt}{0.400pt}}
\multiput(736.00,407.60)(1.651,0.468){5}{\rule{1.300pt}{0.113pt}}
\multiput(736.00,406.17)(9.302,4.000){2}{\rule{0.650pt}{0.400pt}}
\multiput(748.00,411.60)(1.651,0.468){5}{\rule{1.300pt}{0.113pt}}
\multiput(748.00,410.17)(9.302,4.000){2}{\rule{0.650pt}{0.400pt}}
\multiput(760.00,415.60)(1.797,0.468){5}{\rule{1.400pt}{0.113pt}}
\multiput(760.00,414.17)(10.094,4.000){2}{\rule{0.700pt}{0.400pt}}
\multiput(773.00,419.59)(1.267,0.477){7}{\rule{1.060pt}{0.115pt}}
\multiput(773.00,418.17)(9.800,5.000){2}{\rule{0.530pt}{0.400pt}}
\multiput(785.00,424.60)(1.651,0.468){5}{\rule{1.300pt}{0.113pt}}
\multiput(785.00,423.17)(9.302,4.000){2}{\rule{0.650pt}{0.400pt}}
\multiput(797.00,428.60)(1.797,0.468){5}{\rule{1.400pt}{0.113pt}}
\multiput(797.00,427.17)(10.094,4.000){2}{\rule{0.700pt}{0.400pt}}
\multiput(810.00,432.60)(1.651,0.468){5}{\rule{1.300pt}{0.113pt}}
\multiput(810.00,431.17)(9.302,4.000){2}{\rule{0.650pt}{0.400pt}}
\multiput(822.00,436.60)(1.651,0.468){5}{\rule{1.300pt}{0.113pt}}
\multiput(822.00,435.17)(9.302,4.000){2}{\rule{0.650pt}{0.400pt}}
\multiput(834.00,440.60)(1.651,0.468){5}{\rule{1.300pt}{0.113pt}}
\multiput(834.00,439.17)(9.302,4.000){2}{\rule{0.650pt}{0.400pt}}
\multiput(846.00,444.59)(1.378,0.477){7}{\rule{1.140pt}{0.115pt}}
\multiput(846.00,443.17)(10.634,5.000){2}{\rule{0.570pt}{0.400pt}}
\multiput(859.00,449.60)(1.651,0.468){5}{\rule{1.300pt}{0.113pt}}
\multiput(859.00,448.17)(9.302,4.000){2}{\rule{0.650pt}{0.400pt}}
\multiput(871.00,453.60)(1.651,0.468){5}{\rule{1.300pt}{0.113pt}}
\multiput(871.00,452.17)(9.302,4.000){2}{\rule{0.650pt}{0.400pt}}
\multiput(883.00,457.60)(1.797,0.468){5}{\rule{1.400pt}{0.113pt}}
\multiput(883.00,456.17)(10.094,4.000){2}{\rule{0.700pt}{0.400pt}}
\multiput(896.00,461.60)(1.651,0.468){5}{\rule{1.300pt}{0.113pt}}
\multiput(896.00,460.17)(9.302,4.000){2}{\rule{0.650pt}{0.400pt}}
\multiput(908.00,465.60)(1.651,0.468){5}{\rule{1.300pt}{0.113pt}}
\multiput(908.00,464.17)(9.302,4.000){2}{\rule{0.650pt}{0.400pt}}
\multiput(920.00,469.59)(1.267,0.477){7}{\rule{1.060pt}{0.115pt}}
\multiput(920.00,468.17)(9.800,5.000){2}{\rule{0.530pt}{0.400pt}}
\multiput(932.00,474.60)(1.797,0.468){5}{\rule{1.400pt}{0.113pt}}
\multiput(932.00,473.17)(10.094,4.000){2}{\rule{0.700pt}{0.400pt}}
\multiput(945.00,478.60)(1.651,0.468){5}{\rule{1.300pt}{0.113pt}}
\multiput(945.00,477.17)(9.302,4.000){2}{\rule{0.650pt}{0.400pt}}
\multiput(957.00,482.60)(1.651,0.468){5}{\rule{1.300pt}{0.113pt}}
\multiput(957.00,481.17)(9.302,4.000){2}{\rule{0.650pt}{0.400pt}}
\multiput(969.00,486.60)(1.797,0.468){5}{\rule{1.400pt}{0.113pt}}
\multiput(969.00,485.17)(10.094,4.000){2}{\rule{0.700pt}{0.400pt}}
\multiput(982.00,490.60)(1.651,0.468){5}{\rule{1.300pt}{0.113pt}}
\multiput(982.00,489.17)(9.302,4.000){2}{\rule{0.650pt}{0.400pt}}
\multiput(994.00,494.59)(1.267,0.477){7}{\rule{1.060pt}{0.115pt}}
\multiput(994.00,493.17)(9.800,5.000){2}{\rule{0.530pt}{0.400pt}}
\multiput(1006.00,499.60)(1.651,0.468){5}{\rule{1.300pt}{0.113pt}}
\multiput(1006.00,498.17)(9.302,4.000){2}{\rule{0.650pt}{0.400pt}}
\multiput(1018.00,503.60)(1.797,0.468){5}{\rule{1.400pt}{0.113pt}}
\multiput(1018.00,502.17)(10.094,4.000){2}{\rule{0.700pt}{0.400pt}}
\multiput(1031.00,507.60)(1.651,0.468){5}{\rule{1.300pt}{0.113pt}}
\multiput(1031.00,506.17)(9.302,4.000){2}{\rule{0.650pt}{0.400pt}}
\multiput(1043.00,511.60)(1.651,0.468){5}{\rule{1.300pt}{0.113pt}}
\multiput(1043.00,510.17)(9.302,4.000){2}{\rule{0.650pt}{0.400pt}}
\multiput(1055.00,515.60)(1.797,0.468){5}{\rule{1.400pt}{0.113pt}}
\multiput(1055.00,514.17)(10.094,4.000){2}{\rule{0.700pt}{0.400pt}}
\multiput(1068.00,519.59)(1.267,0.477){7}{\rule{1.060pt}{0.115pt}}
\multiput(1068.00,518.17)(9.800,5.000){2}{\rule{0.530pt}{0.400pt}}
\multiput(1080.00,524.60)(1.651,0.468){5}{\rule{1.300pt}{0.113pt}}
\multiput(1080.00,523.17)(9.302,4.000){2}{\rule{0.650pt}{0.400pt}}
\multiput(1092.00,528.60)(1.651,0.468){5}{\rule{1.300pt}{0.113pt}}
\multiput(1092.00,527.17)(9.302,4.000){2}{\rule{0.650pt}{0.400pt}}
\multiput(1104.00,532.60)(1.797,0.468){5}{\rule{1.400pt}{0.113pt}}
\multiput(1104.00,531.17)(10.094,4.000){2}{\rule{0.700pt}{0.400pt}}
\multiput(1117.00,536.60)(1.651,0.468){5}{\rule{1.300pt}{0.113pt}}
\multiput(1117.00,535.17)(9.302,4.000){2}{\rule{0.650pt}{0.400pt}}
\multiput(1129.00,540.60)(1.651,0.468){5}{\rule{1.300pt}{0.113pt}}
\multiput(1129.00,539.17)(9.302,4.000){2}{\rule{0.650pt}{0.400pt}}
\multiput(1141.00,544.59)(1.267,0.477){7}{\rule{1.060pt}{0.115pt}}
\multiput(1141.00,543.17)(9.800,5.000){2}{\rule{0.530pt}{0.400pt}}
\multiput(1153.00,549.60)(1.797,0.468){5}{\rule{1.400pt}{0.113pt}}
\multiput(1153.00,548.17)(10.094,4.000){2}{\rule{0.700pt}{0.400pt}}
\multiput(1166.00,553.60)(1.651,0.468){5}{\rule{1.300pt}{0.113pt}}
\multiput(1166.00,552.17)(9.302,4.000){2}{\rule{0.650pt}{0.400pt}}
\multiput(1178.00,557.60)(1.651,0.468){5}{\rule{1.300pt}{0.113pt}}
\multiput(1178.00,556.17)(9.302,4.000){2}{\rule{0.650pt}{0.400pt}}
\multiput(1190.00,561.60)(1.797,0.468){5}{\rule{1.400pt}{0.113pt}}
\multiput(1190.00,560.17)(10.094,4.000){2}{\rule{0.700pt}{0.400pt}}
\multiput(1203.00,565.60)(1.651,0.468){5}{\rule{1.300pt}{0.113pt}}
\multiput(1203.00,564.17)(9.302,4.000){2}{\rule{0.650pt}{0.400pt}}
\multiput(1215.00,569.59)(1.267,0.477){7}{\rule{1.060pt}{0.115pt}}
\multiput(1215.00,568.17)(9.800,5.000){2}{\rule{0.530pt}{0.400pt}}
\multiput(1227.00,574.60)(1.651,0.468){5}{\rule{1.300pt}{0.113pt}}
\multiput(1227.00,573.17)(9.302,4.000){2}{\rule{0.650pt}{0.400pt}}
\multiput(1239.00,578.60)(1.797,0.468){5}{\rule{1.400pt}{0.113pt}}
\multiput(1239.00,577.17)(10.094,4.000){2}{\rule{0.700pt}{0.400pt}}
\multiput(1252.00,582.60)(1.651,0.468){5}{\rule{1.300pt}{0.113pt}}
\multiput(1252.00,581.17)(9.302,4.000){2}{\rule{0.650pt}{0.400pt}}
\multiput(1264.00,586.60)(1.651,0.468){5}{\rule{1.300pt}{0.113pt}}
\multiput(1264.00,585.17)(9.302,4.000){2}{\rule{0.650pt}{0.400pt}}
\multiput(1276.00,590.60)(1.797,0.468){5}{\rule{1.400pt}{0.113pt}}
\multiput(1276.00,589.17)(10.094,4.000){2}{\rule{0.700pt}{0.400pt}}
\multiput(1289.00,594.59)(1.267,0.477){7}{\rule{1.060pt}{0.115pt}}
\multiput(1289.00,593.17)(9.800,5.000){2}{\rule{0.530pt}{0.400pt}}
\multiput(1301.00,599.60)(1.651,0.468){5}{\rule{1.300pt}{0.113pt}}
\multiput(1301.00,598.17)(9.302,4.000){2}{\rule{0.650pt}{0.400pt}}
\multiput(1313.00,603.60)(1.651,0.468){5}{\rule{1.300pt}{0.113pt}}
\multiput(1313.00,602.17)(9.302,4.000){2}{\rule{0.650pt}{0.400pt}}
\multiput(1325.00,607.60)(1.797,0.468){5}{\rule{1.400pt}{0.113pt}}
\multiput(1325.00,606.17)(10.094,4.000){2}{\rule{0.700pt}{0.400pt}}
\multiput(1338.00,611.60)(1.651,0.468){5}{\rule{1.300pt}{0.113pt}}
\multiput(1338.00,610.17)(9.302,4.000){2}{\rule{0.650pt}{0.400pt}}
\multiput(1350.00,615.60)(1.651,0.468){5}{\rule{1.300pt}{0.113pt}}
\multiput(1350.00,614.17)(9.302,4.000){2}{\rule{0.650pt}{0.400pt}}
\multiput(1362.00,619.59)(1.378,0.477){7}{\rule{1.140pt}{0.115pt}}
\multiput(1362.00,618.17)(10.634,5.000){2}{\rule{0.570pt}{0.400pt}}
\multiput(1375.00,624.60)(1.651,0.468){5}{\rule{1.300pt}{0.113pt}}
\multiput(1375.00,623.17)(9.302,4.000){2}{\rule{0.650pt}{0.400pt}}
\multiput(1387.00,628.60)(1.651,0.468){5}{\rule{1.300pt}{0.113pt}}
\multiput(1387.00,627.17)(9.302,4.000){2}{\rule{0.650pt}{0.400pt}}
\multiput(1399.00,632.60)(1.651,0.468){5}{\rule{1.300pt}{0.113pt}}
\multiput(1399.00,631.17)(9.302,4.000){2}{\rule{0.650pt}{0.400pt}}
\multiput(1411.00,636.60)(1.797,0.468){5}{\rule{1.400pt}{0.113pt}}
\multiput(1411.00,635.17)(10.094,4.000){2}{\rule{0.700pt}{0.400pt}}
\multiput(1424.00,640.60)(1.651,0.468){5}{\rule{1.300pt}{0.113pt}}
\multiput(1424.00,639.17)(9.302,4.000){2}{\rule{0.650pt}{0.400pt}}
\put(539,113){\usebox{\plotpoint}}
\put(539.00,113.00){\usebox{\plotpoint}}
\put(554.17,127.17){\usebox{\plotpoint}}
\put(568.65,142.03){\usebox{\plotpoint}}
\put(583.01,157.01){\usebox{\plotpoint}}
\put(598.06,171.29){\usebox{\plotpoint}}
\put(612.85,185.85){\usebox{\plotpoint}}
\multiput(613,186)(14.676,14.676){0}{\usebox{\plotpoint}}
\put(627.63,200.43){\usebox{\plotpoint}}
\put(642.69,214.69){\usebox{\plotpoint}}
\put(657.37,229.37){\usebox{\plotpoint}}
\put(672.05,244.05){\usebox{\plotpoint}}
\multiput(674,246)(15.251,14.078){0}{\usebox{\plotpoint}}
\put(687.21,258.21){\usebox{\plotpoint}}
\put(701.89,272.89){\usebox{\plotpoint}}
\put(716.78,287.34){\usebox{\plotpoint}}
\put(731.73,301.73){\usebox{\plotpoint}}
\put(746.41,316.41){\usebox{\plotpoint}}
\multiput(748,318)(14.676,14.676){0}{\usebox{\plotpoint}}
\put(761.13,331.04){\usebox{\plotpoint}}
\put(776.25,345.25){\usebox{\plotpoint}}
\put(790.93,359.93){\usebox{\plotpoint}}
\put(805.94,374.25){\usebox{\plotpoint}}
\put(820.77,388.77){\usebox{\plotpoint}}
\multiput(822,390)(14.676,14.676){0}{\usebox{\plotpoint}}
\put(835.45,403.45){\usebox{\plotpoint}}
\put(850.28,417.96){\usebox{\plotpoint}}
\put(865.29,432.29){\usebox{\plotpoint}}
\put(879.60,447.32){\usebox{\plotpoint}}
\put(894.57,461.68){\usebox{\plotpoint}}
\multiput(896,463)(14.676,14.676){0}{\usebox{\plotpoint}}
\put(909.30,476.30){\usebox{\plotpoint}}
\put(923.98,490.98){\usebox{\plotpoint}}
\put(938.91,505.38){\usebox{\plotpoint}}
\put(953.82,519.82){\usebox{\plotpoint}}
\put(968.49,534.49){\usebox{\plotpoint}}
\multiput(969,535)(15.251,14.078){0}{\usebox{\plotpoint}}
\put(983.66,548.66){\usebox{\plotpoint}}
\put(998.34,563.34){\usebox{\plotpoint}}
\put(1013.01,578.01){\usebox{\plotpoint}}
\put(1028.07,592.29){\usebox{\plotpoint}}
\put(1042.86,606.86){\usebox{\plotpoint}}
\multiput(1043,607)(14.676,14.676){0}{\usebox{\plotpoint}}
\put(1057.63,621.43){\usebox{\plotpoint}}
\put(1072.70,635.70){\usebox{\plotpoint}}
\put(1087.38,650.38){\usebox{\plotpoint}}
\put(1102.05,665.05){\usebox{\plotpoint}}
\multiput(1104,667)(15.251,14.078){0}{\usebox{\plotpoint}}
\put(1117.22,679.22){\usebox{\plotpoint}}
\put(1131.89,693.89){\usebox{\plotpoint}}
\put(1146.57,708.57){\usebox{\plotpoint}}
\put(1161.57,722.91){\usebox{\plotpoint}}
\put(1176.41,737.41){\usebox{\plotpoint}}
\multiput(1178,739)(14.676,14.676){0}{\usebox{\plotpoint}}
\put(1191.09,752.09){\usebox{\plotpoint}}
\put(1205.77,766.77){\usebox{\plotpoint}}
\put(1220.44,781.44){\usebox{\plotpoint}}
\put(1235.12,796.12){\usebox{\plotpoint}}
\put(1250.22,810.36){\usebox{\plotpoint}}
\multiput(1252,812)(14.676,14.676){0}{\usebox{\plotpoint}}
\put(1264.96,824.96){\usebox{\plotpoint}}
\put(1279.78,839.49){\usebox{\plotpoint}}
\put(1294.80,853.80){\usebox{\plotpoint}}
\put(1309.48,868.48){\usebox{\plotpoint}}
\multiput(1313,872)(14.676,14.676){0}{\usebox{\plotpoint}}
\put(1318,877){\usebox{\plotpoint}}
\sbox{\plotpoint}{\rule[-0.400pt]{0.800pt}{0.800pt}}%
\put(588,321){\raisebox{-.8pt}{\makebox(0,0){$\Diamond$}}}
\put(600,329){\raisebox{-.8pt}{\makebox(0,0){$\Diamond$}}}
\put(612,332){\raisebox{-.8pt}{\makebox(0,0){$\Diamond$}}}
\put(625,341){\raisebox{-.8pt}{\makebox(0,0){$\Diamond$}}}
\put(639,351){\raisebox{-.8pt}{\makebox(0,0){$\Diamond$}}}
\put(654,370){\raisebox{-.8pt}{\makebox(0,0){$\Diamond$}}}
\put(670,379){\raisebox{-.8pt}{\makebox(0,0){$\Diamond$}}}
\put(688,388){\raisebox{-.8pt}{\makebox(0,0){$\Diamond$}}}
\put(706,393){\raisebox{-.8pt}{\makebox(0,0){$\Diamond$}}}
\put(727,409){\raisebox{-.8pt}{\makebox(0,0){$\Diamond$}}}
\put(749,430){\raisebox{-.8pt}{\makebox(0,0){$\Diamond$}}}
\put(773,450){\raisebox{-.8pt}{\makebox(0,0){$\Diamond$}}}
\put(799,463){\raisebox{-.8pt}{\makebox(0,0){$\Diamond$}}}
\put(828,489){\raisebox{-.8pt}{\makebox(0,0){$\Diamond$}}}
\put(860,521){\raisebox{-.8pt}{\makebox(0,0){$\Diamond$}}}
\put(896,541){\raisebox{-.8pt}{\makebox(0,0){$\Diamond$}}}
\put(935,577){\raisebox{-.8pt}{\makebox(0,0){$\Diamond$}}}
\put(980,626){\raisebox{-.8pt}{\makebox(0,0){$\Diamond$}}}
\put(1031,644){\raisebox{-.8pt}{\makebox(0,0){$\Diamond$}}}
\put(1089,700){\raisebox{-.8pt}{\makebox(0,0){$\Diamond$}}}
\put(1155,712){\raisebox{-.8pt}{\makebox(0,0){$\Diamond$}}}
\put(1233,811){\raisebox{-.8pt}{\makebox(0,0){$\Diamond$}}}
\sbox{\plotpoint}{\rule[-0.500pt]{1.000pt}{1.000pt}}%
\put(588,318){\makebox(0,0){$+$}}
\put(600,327){\makebox(0,0){$+$}}
\put(612,332){\makebox(0,0){$+$}}
\put(625,343){\makebox(0,0){$+$}}
\put(639,349){\makebox(0,0){$+$}}
\put(654,354){\makebox(0,0){$+$}}
\put(670,364){\makebox(0,0){$+$}}
\put(688,375){\makebox(0,0){$+$}}
\put(706,388){\makebox(0,0){$+$}}
\put(727,396){\makebox(0,0){$+$}}
\put(749,403){\makebox(0,0){$+$}}
\put(773,417){\makebox(0,0){$+$}}
\put(799,433){\makebox(0,0){$+$}}
\put(828,443){\makebox(0,0){$+$}}
\put(860,471){\makebox(0,0){$+$}}
\put(896,480){\makebox(0,0){$+$}}
\put(935,513){\makebox(0,0){$+$}}
\put(980,546){\makebox(0,0){$+$}}
\put(1031,590){\makebox(0,0){$+$}}
\put(1089,650){\makebox(0,0){$+$}}
\put(1155,694){\makebox(0,0){$+$}}
\put(1233,751){\makebox(0,0){$+$}}
\sbox{\plotpoint}{\rule[-0.600pt]{1.200pt}{1.200pt}}%
\put(588,321){\raisebox{-.8pt}{\makebox(0,0){$\Box$}}}
\put(600,324){\raisebox{-.8pt}{\makebox(0,0){$\Box$}}}
\put(612,334){\raisebox{-.8pt}{\makebox(0,0){$\Box$}}}
\put(625,339){\raisebox{-.8pt}{\makebox(0,0){$\Box$}}}
\put(639,349){\raisebox{-.8pt}{\makebox(0,0){$\Box$}}}
\put(654,356){\raisebox{-.8pt}{\makebox(0,0){$\Box$}}}
\put(670,364){\raisebox{-.8pt}{\makebox(0,0){$\Box$}}}
\put(688,375){\raisebox{-.8pt}{\makebox(0,0){$\Box$}}}
\put(706,381){\raisebox{-.8pt}{\makebox(0,0){$\Box$}}}
\put(727,393){\raisebox{-.8pt}{\makebox(0,0){$\Box$}}}
\put(749,400){\raisebox{-.8pt}{\makebox(0,0){$\Box$}}}
\put(773,412){\raisebox{-.8pt}{\makebox(0,0){$\Box$}}}
\put(799,427){\raisebox{-.8pt}{\makebox(0,0){$\Box$}}}
\put(828,442){\raisebox{-.8pt}{\makebox(0,0){$\Box$}}}
\put(860,454){\raisebox{-.8pt}{\makebox(0,0){$\Box$}}}
\put(896,471){\raisebox{-.8pt}{\makebox(0,0){$\Box$}}}
\put(935,495){\raisebox{-.8pt}{\makebox(0,0){$\Box$}}}
\put(980,514){\raisebox{-.8pt}{\makebox(0,0){$\Box$}}}
\put(1031,547){\raisebox{-.8pt}{\makebox(0,0){$\Box$}}}
\put(1089,582){\raisebox{-.8pt}{\makebox(0,0){$\Box$}}}
\put(1155,621){\raisebox{-.8pt}{\makebox(0,0){$\Box$}}}
\put(1233,681){\raisebox{-.8pt}{\makebox(0,0){$\Box$}}}
\sbox{\plotpoint}{\rule[-0.500pt]{1.000pt}{1.000pt}}%
\put(588,318){\makebox(0,0){$\times$}}
\put(600,324){\makebox(0,0){$\times$}}
\put(612,334){\makebox(0,0){$\times$}}
\put(625,339){\makebox(0,0){$\times$}}
\put(639,347){\makebox(0,0){$\times$}}
\put(654,356){\makebox(0,0){$\times$}}
\put(670,366){\makebox(0,0){$\times$}}
\put(688,373){\makebox(0,0){$\times$}}
\put(706,381){\makebox(0,0){$\times$}}
\put(727,392){\makebox(0,0){$\times$}}
\put(749,402){\makebox(0,0){$\times$}}
\put(773,414){\makebox(0,0){$\times$}}
\put(799,427){\makebox(0,0){$\times$}}
\put(828,437){\makebox(0,0){$\times$}}
\put(860,451){\makebox(0,0){$\times$}}
\put(896,464){\makebox(0,0){$\times$}}
\put(935,481){\makebox(0,0){$\times$}}
\put(980,501){\makebox(0,0){$\times$}}
\put(1031,521){\makebox(0,0){$\times$}}
\put(1089,549){\makebox(0,0){$\times$}}
\put(1155,583){\makebox(0,0){$\times$}}
\put(1233,654){\makebox(0,0){$\times$}}
\put(1325,685){\makebox(0,0){$\times$}}
\sbox{\plotpoint}{\rule[-0.200pt]{0.400pt}{0.400pt}}%
\put(588,318){\makebox(0,0){$\triangle$}}
\put(600,327){\makebox(0,0){$\triangle$}}
\put(612,332){\makebox(0,0){$\triangle$}}
\put(625,341){\makebox(0,0){$\triangle$}}
\put(639,347){\makebox(0,0){$\triangle$}}
\put(654,354){\makebox(0,0){$\triangle$}}
\put(670,364){\makebox(0,0){$\triangle$}}
\put(688,371){\makebox(0,0){$\triangle$}}
\put(706,382){\makebox(0,0){$\triangle$}}
\put(727,393){\makebox(0,0){$\triangle$}}
\put(749,403){\makebox(0,0){$\triangle$}}
\put(773,414){\makebox(0,0){$\triangle$}}
\put(799,426){\makebox(0,0){$\triangle$}}
\put(828,438){\makebox(0,0){$\triangle$}}
\put(860,451){\makebox(0,0){$\triangle$}}
\put(896,464){\makebox(0,0){$\triangle$}}
\put(935,482){\makebox(0,0){$\triangle$}}
\put(980,496){\makebox(0,0){$\triangle$}}
\put(1031,517){\makebox(0,0){$\triangle$}}
\put(1089,546){\makebox(0,0){$\triangle$}}
\put(1155,586){\makebox(0,0){$\triangle$}}
\put(1233,614){\makebox(0,0){$\triangle$}}
\put(588,318){\makebox(0,0){$\star$}}
\put(600,324){\makebox(0,0){$\star$}}
\put(612,332){\makebox(0,0){$\star$}}
\put(625,339){\makebox(0,0){$\star$}}
\put(639,345){\makebox(0,0){$\star$}}
\put(654,354){\makebox(0,0){$\star$}}
\put(670,363){\makebox(0,0){$\star$}}
\put(688,374){\makebox(0,0){$\star$}}
\put(706,382){\makebox(0,0){$\star$}}
\put(727,390){\makebox(0,0){$\star$}}
\put(749,401){\makebox(0,0){$\star$}}
\put(773,414){\makebox(0,0){$\star$}}
\put(799,424){\makebox(0,0){$\star$}}
\put(828,437){\makebox(0,0){$\star$}}
\put(860,450){\makebox(0,0){$\star$}}
\put(896,465){\makebox(0,0){$\star$}}
\put(935,476){\makebox(0,0){$\star$}}
\put(980,492){\makebox(0,0){$\star$}}
\put(1031,512){\makebox(0,0){$\star$}}
\put(1089,532){\makebox(0,0){$\star$}}
\put(1155,569){\makebox(0,0){$\star$}}
\put(1233,593){\makebox(0,0){$\star$}}
\sbox{\plotpoint}{\rule[-0.400pt]{0.800pt}{0.800pt}}%
\put(588,318){\circle{12}}
\put(600,324){\circle{12}}
\put(612,332){\circle{12}}
\put(625,339){\circle{12}}
\put(639,347){\circle{12}}
\put(654,354){\circle{12}}
\put(670,364){\circle{12}}
\put(688,373){\circle{12}}
\put(706,381){\circle{12}}
\put(727,391){\circle{12}}
\put(749,401){\circle{12}}
\put(773,412){\circle{12}}
\put(799,425){\circle{12}}
\put(828,437){\circle{12}}
\put(860,448){\circle{12}}
\put(896,463){\circle{12}}
\put(935,480){\circle{12}}
\put(980,494){\circle{12}}
\put(1031,515){\circle{12}}
\put(1089,532){\circle{12}}
\put(1155,552){\circle{12}}
\put(1233,594){\circle{12}}
\end{picture}
\bigskip

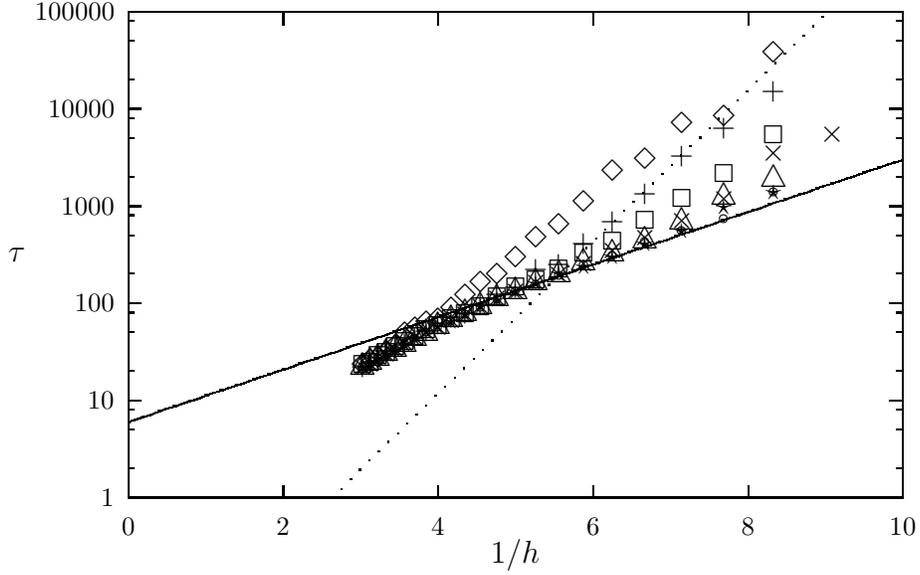
\noindent Fig. 1. Nucleation time ($\tau$) is plotted against the inverse field
($1/h$) for various system sizes. ($\Diamond$) L = 51, ($+$) L = 101,
($\Box$) L = 151, ($\times$) L = 201, ($\triangle$) L = 251, ($\star$)
 L = 351 and (~{\circle{26}}) L =401. $J/KT$ = 5/8 here. The two straight lines
are the best fitted straight lines in the coalescence and nucleation regimes.

\newpage
%FIG 2
\setlength{\unitlength}{0.240900pt}
\ifx\plotpoint\undefined\newsavebox{\plotpoint}\fi
\sbox{\plotpoint}{\rule[-0.200pt]{0.400pt}{0.400pt}}%
\begin{picture}(1500,900)(0,0)
\font\gnuplot=cmr10 at 10pt
\gnuplot
\sbox{\plotpoint}{\rule[-0.200pt]{0.400pt}{0.400pt}}%
\put(220.0,113.0){\rule[-0.200pt]{4.818pt}{0.400pt}}
\put(198,113){\makebox(0,0)[r]{0.12}}
\put(1416.0,113.0){\rule[-0.200pt]{4.818pt}{0.400pt}}
\put(220.0,240.0){\rule[-0.200pt]{4.818pt}{0.400pt}}
\put(198,240){\makebox(0,0)[r]{0.14}}
\put(1416.0,240.0){\rule[-0.200pt]{4.818pt}{0.400pt}}
\put(220.0,368.0){\rule[-0.200pt]{4.818pt}{0.400pt}}
\put(198,368){\makebox(0,0)[r]{0.16}}
\put(1416.0,368.0){\rule[-0.200pt]{4.818pt}{0.400pt}}
\put(220.0,495.0){\rule[-0.200pt]{4.818pt}{0.400pt}}
\put(198,495){\makebox(0,0)[r]{0.18}}
\put(1416.0,495.0){\rule[-0.200pt]{4.818pt}{0.400pt}}
\put(220.0,622.0){\rule[-0.200pt]{4.818pt}{0.400pt}}
\put(198,622){\makebox(0,0)[r]{0.2}}
\put(1416.0,622.0){\rule[-0.200pt]{4.818pt}{0.400pt}}
\put(220.0,750.0){\rule[-0.200pt]{4.818pt}{0.400pt}}
\put(198,750){\makebox(0,0)[r]{0.22}}
\put(1416.0,750.0){\rule[-0.200pt]{4.818pt}{0.400pt}}
\put(220.0,877.0){\rule[-0.200pt]{4.818pt}{0.400pt}}
\put(198,877){\makebox(0,0)[r]{0.24}}
\put(1416.0,877.0){\rule[-0.200pt]{4.818pt}{0.400pt}}
\put(220.0,113.0){\rule[-0.200pt]{0.400pt}{4.818pt}}
\put(220,68){\makebox(0,0){0.18}}
\put(220.0,857.0){\rule[-0.200pt]{0.400pt}{4.818pt}}
\put(372.0,113.0){\rule[-0.200pt]{0.400pt}{4.818pt}}
\put(372,68){\makebox(0,0){0.19}}
\put(372.0,857.0){\rule[-0.200pt]{0.400pt}{4.818pt}}
\put(524.0,113.0){\rule[-0.200pt]{0.400pt}{4.818pt}}
\put(524,68){\makebox(0,0){0.2}}
\put(524.0,857.0){\rule[-0.200pt]{0.400pt}{4.818pt}}
\put(676.0,113.0){\rule[-0.200pt]{0.400pt}{4.818pt}}
\put(676,68){\makebox(0,0){0.21}}
\put(676.0,857.0){\rule[-0.200pt]{0.400pt}{4.818pt}}
\put(828.0,113.0){\rule[-0.200pt]{0.400pt}{4.818pt}}
\put(828,68){\makebox(0,0){0.22}}
\put(828.0,857.0){\rule[-0.200pt]{0.400pt}{4.818pt}}
\put(980.0,113.0){\rule[-0.200pt]{0.400pt}{4.818pt}}
\put(980,68){\makebox(0,0){0.23}}
\put(980.0,857.0){\rule[-0.200pt]{0.400pt}{4.818pt}}
\put(1132.0,113.0){\rule[-0.200pt]{0.400pt}{4.818pt}}
\put(1132,68){\makebox(0,0){0.24}}
\put(1132.0,857.0){\rule[-0.200pt]{0.400pt}{4.818pt}}
\put(1284.0,113.0){\rule[-0.200pt]{0.400pt}{4.818pt}}
\put(1284,68){\makebox(0,0){0.25}}
\put(1284.0,857.0){\rule[-0.200pt]{0.400pt}{4.818pt}}
\put(1436.0,113.0){\rule[-0.200pt]{0.400pt}{4.818pt}}
\put(1436,68){\makebox(0,0){0.26}}
\put(1436.0,857.0){\rule[-0.200pt]{0.400pt}{4.818pt}}
\put(220.0,113.0){\rule[-0.200pt]{292.934pt}{0.400pt}}
\put(1436.0,113.0){\rule[-0.200pt]{0.400pt}{184.048pt}}
\put(220.0,877.0){\rule[-0.200pt]{292.934pt}{0.400pt}}
\put(45,495){\makebox(0,0){$h_c$}}
\put(828,23){\makebox(0,0){$1/\log(L)$}}
\put(220.0,113.0){\rule[-0.200pt]{0.400pt}{184.048pt}}
\put(1349,877){\raisebox{-.8pt}{\makebox(0,0){$\Diamond$}}}
\put(776,495){\raisebox{-.8pt}{\makebox(0,0){$\Diamond$}}}
\put(513,368){\raisebox{-.8pt}{\makebox(0,0){$\Diamond$}}}
\put(349,240){\raisebox{-.8pt}{\makebox(0,0){$\Diamond$}}}
\put(234,177){\raisebox{-.8pt}{\makebox(0,0){$\Diamond$}}}
\end{picture}

\bigskip

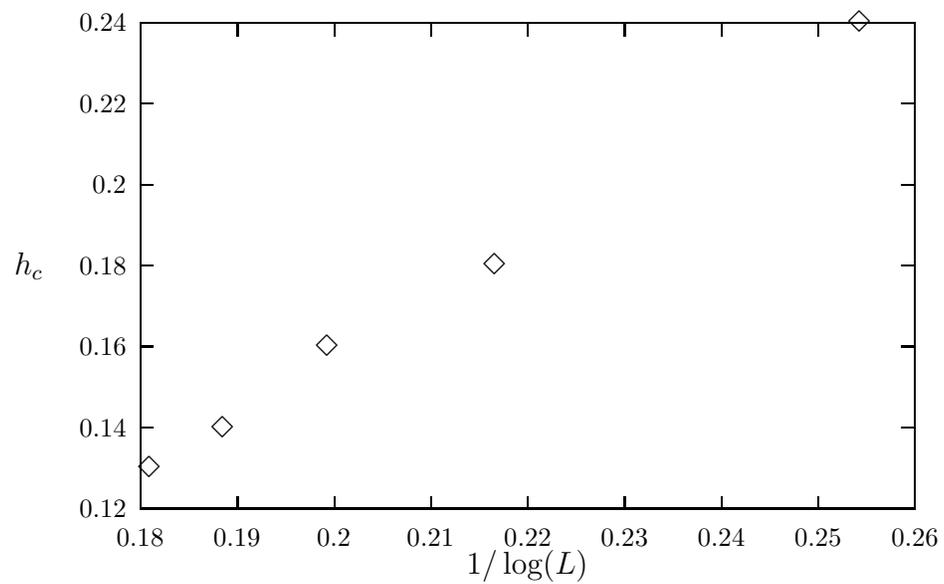
\centerline {Fig.2. The crossover field ($h_c$) is plotted against $1/\log(L)$.}

\newpage
% FIG 3
\setlength{\unitlength}{0.240900pt}
\ifx\plotpoint\undefined\newsavebox{\plotpoint}\fi
\sbox{\plotpoint}{\rule[-0.200pt]{0.400pt}{0.400pt}}%
\begin{picture}(1500,900)(0,0)
\font\gnuplot=cmr10 at 10pt
\gnuplot
\sbox{\plotpoint}{\rule[-0.200pt]{0.400pt}{0.400pt}}%
\put(220.0,113.0){\rule[-0.200pt]{292.934pt}{0.400pt}}
\put(220.0,113.0){\rule[-0.200pt]{0.400pt}{184.048pt}}
\put(220.0,113.0){\rule[-0.200pt]{4.818pt}{0.400pt}}
\put(198,113){\makebox(0,0)[r]{0}}
\put(1416.0,113.0){\rule[-0.200pt]{4.818pt}{0.400pt}}
\put(220.0,222.0){\rule[-0.200pt]{4.818pt}{0.400pt}}
\put(198,222){\makebox(0,0)[r]{200}}
\put(1416.0,222.0){\rule[-0.200pt]{4.818pt}{0.400pt}}
\put(220.0,331.0){\rule[-0.200pt]{4.818pt}{0.400pt}}
\put(198,331){\makebox(0,0)[r]{400}}
\put(1416.0,331.0){\rule[-0.200pt]{4.818pt}{0.400pt}}
\put(220.0,440.0){\rule[-0.200pt]{4.818pt}{0.400pt}}
\put(198,440){\makebox(0,0)[r]{600}}
\put(1416.0,440.0){\rule[-0.200pt]{4.818pt}{0.400pt}}
\put(220.0,550.0){\rule[-0.200pt]{4.818pt}{0.400pt}}
\put(198,550){\makebox(0,0)[r]{800}}
\put(1416.0,550.0){\rule[-0.200pt]{4.818pt}{0.400pt}}
\put(220.0,659.0){\rule[-0.200pt]{4.818pt}{0.400pt}}
\put(198,659){\makebox(0,0)[r]{1000}}
\put(1416.0,659.0){\rule[-0.200pt]{4.818pt}{0.400pt}}
\put(220.0,768.0){\rule[-0.200pt]{4.818pt}{0.400pt}}
\put(198,768){\makebox(0,0)[r]{1200}}
\put(1416.0,768.0){\rule[-0.200pt]{4.818pt}{0.400pt}}
\put(220.0,877.0){\rule[-0.200pt]{4.818pt}{0.400pt}}
\put(198,877){\makebox(0,0)[r]{1400}}
\put(1416.0,877.0){\rule[-0.200pt]{4.818pt}{0.400pt}}
\put(220.0,113.0){\rule[-0.200pt]{0.400pt}{4.818pt}}
\put(220,68){\makebox(0,0){0}}
\put(220.0,857.0){\rule[-0.200pt]{0.400pt}{4.818pt}}
\put(394.0,113.0){\rule[-0.200pt]{0.400pt}{4.818pt}}
\put(394,68){\makebox(0,0){10000}}
\put(394.0,857.0){\rule[-0.200pt]{0.400pt}{4.818pt}}
\put(567.0,113.0){\rule[-0.200pt]{0.400pt}{4.818pt}}
\put(567,68){\makebox(0,0){20000}}
\put(567.0,857.0){\rule[-0.200pt]{0.400pt}{4.818pt}}
\put(741.0,113.0){\rule[-0.200pt]{0.400pt}{4.818pt}}
\put(741,68){\makebox(0,0){30000}}
\put(741.0,857.0){\rule[-0.200pt]{0.400pt}{4.818pt}}
\put(915.0,113.0){\rule[-0.200pt]{0.400pt}{4.818pt}}
\put(915,68){\makebox(0,0){40000}}
\put(915.0,857.0){\rule[-0.200pt]{0.400pt}{4.818pt}}
\put(1089.0,113.0){\rule[-0.200pt]{0.400pt}{4.818pt}}
\put(1089,68){\makebox(0,0){50000}}
\put(1089.0,857.0){\rule[-0.200pt]{0.400pt}{4.818pt}}
\put(1262.0,113.0){\rule[-0.200pt]{0.400pt}{4.818pt}}
\put(1262,68){\makebox(0,0){60000}}
\put(1262.0,857.0){\rule[-0.200pt]{0.400pt}{4.818pt}}
\put(1436.0,113.0){\rule[-0.200pt]{0.400pt}{4.818pt}}
\put(1436,68){\makebox(0,0){70000}}
\put(1436.0,857.0){\rule[-0.200pt]{0.400pt}{4.818pt}}
\put(220.0,113.0){\rule[-0.200pt]{292.934pt}{0.400pt}}
\put(1436.0,113.0){\rule[-0.200pt]{0.400pt}{184.048pt}}
\put(220.0,877.0){\rule[-0.200pt]{292.934pt}{0.400pt}}
\put(45,495){\makebox(0,0){$\tau_c$}}
\put(828,23){\makebox(0,0){$L^2$}}
\put(220.0,113.0){\rule[-0.200pt]{0.400pt}{184.048pt}}
\put(265,160){\raisebox{-.8pt}{\makebox(0,0){$\Diamond$}}}
\put(397,252){\raisebox{-.8pt}{\makebox(0,0){$\Diamond$}}}
\put(616,344){\raisebox{-.8pt}{\makebox(0,0){$\Diamond$}}}
\put(922,503){\raisebox{-.8pt}{\makebox(0,0){$\Diamond$}}}
\put(1314,794){\raisebox{-.8pt}{\makebox(0,0){$\Diamond$}}}
\end{picture}

\bigskip

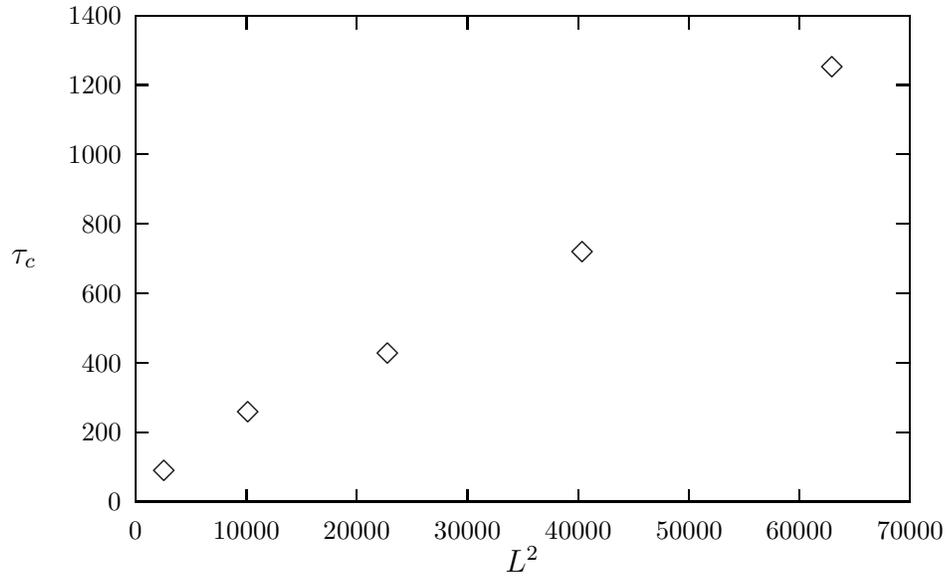
\centerline {Fig. 3. The crossover time ($\tau_c$) is plotted against $L^2$.}
\end{document}